\RequirePackage{fix-cm}
\documentclass[smallextended]{svjour3}       % onecolumn (second format)
\smartqed  % flush right qed marks, e.g. at end of proof
\usepackage{times}
\usepackage{graphicx}
\usepackage{epsf}
\usepackage[colorlinks={true}]{hyperref}

\usepackage[T1]{fontenc}
\usepackage[utf8]{inputenc}
\hypersetup{citecolor={blue}, filecolor={blue}, linkcolor={blue}, urlcolor={blue}}

\newcommand{\commentold}[1]{}
\DeclareMathSymbol{:}{\mathpunct}{operators}{"3A}
\begin{document}

\title{The resource theory of coherence for quantum channels%\thanks{Grants or other notes
}
\subtitle{}


\author{F. H. Kamin$^{1}$\and
        F. T. Tabesh$^{1}$\and S. Salimi$^{1}$\and  F. Kheirandish$^{1}$ %etc.
}


\institute{F. H. Kamin \at
              %Department of Physics, University of Kurdistan, P.O.Box 66177-15175, Sanandaj, Iran \\
              %Tel.: +98-917-9275855\\
              %Fax: +98-71-43567836\\
              \email{f.hatami@sci.uok.ac.ir}           %  \\
           \and
           S. Salimi \at
             %Department of Physics, University of Kurdistan, P.O.Box 66177-15175, Sanandaj, Iran \\
               \email{ShSalimi@uok.ac.ir}
               \and
                $^1$ Department of Physics, University of Kurdistan, P.O.Box 66177-15175, Sanandaj, Iran
}

\date{Received: date / Accepted: date}

\maketitle

\begin{abstract}
We define the quantum-incoherent relative entropy of coherence ($\mathcal{QI}$ REC) of quantum channels in the framework of the resource theory by using the Choi-Jamiolkowsky isomorphism. Coherence-breaking channels are introduced as free operations and their corresponding Choi states as free states. We also show the relationship between the coherence of channel and the quantum discord and find that basis-dependent quantum asymmetric discord can never be more than the $\mathcal{QI}$ REC for any quantum channels. \textbf{Also}, we prove the $\mathcal{QI}$ REC is decreasing for any divisible quantum incoherent channel and we also claim it can be considered as the quantumness of quantum channels. Moreover, we demonstrate that for qubit channels, the relative entropy of coherence (REC) can be equivalent to the REC of their corresponding Choi states and the basis-dependent quantum symmetric discord can never exceed the coherence.
\keywords{Quantum-incoherent relative entropy. Choi states. Coherence-breaking channels. Quantum discord}
\end{abstract}
\section{Introduction}
Quantum resource theories (QRTs) approach play a significant role in quantum information theory, in particular, to determine the measures for evaluating physical resources \cite{r.t1,r.t2,r.t3,r.t4}.  QRTs are defined by constraints that characterize a set of free operations that do not generate a resource and the corresponding set of free states that are devoid of the resource. These restrictions may arise from either fundamental conservation laws or practical restrictions resulting from \textbf{the difficulty} of performing quantum operations. In other words, a strong framework for studying various quantum phenomena can be provided by QRTs. Using this perspective to study quantum systems is natural since their characteristic quantum features can be destroyed due to processes such as decoherence. Taking into account that the structure of QRTs is very general, i.e., the great freedom in \textbf{the definition} of free states and free operations, QRTs can be applied to study many different branches of quantum physics such as entanglement \cite{r.e2,r.e3}, quantum reference frames and asymmetry \cite{as}, quantum thermodynamics \cite{qt}, nonlocality \cite{no}, non-Markovianity \cite{nom}, quantum coherence and superposition \cite{c.a1,c.a2,c.a3,c.a4,c.a5,c.a6,c.a7}, \textbf{etc.} According to \textbf{a common} structure in all QRTs, one can see similarities and \textbf{connections} between different QRTs in terms of available resource measures and resource reversibility \cite{r.t1,co,co1,co2,co3,inco1}. On the other hand, there are cases where multi-resources are required to accomplish a specific task, so attempts have been made to combine the different QRTs as resource theory of thermodynamics that it is a mixture of the purity theory and the asymmetry theory \cite{co3}.

Quantum coherence is one of the most prominent features of quantum systems and it is an important physical resource in many quantum information processes as it can be provided at a specific cost and consumed to accomplish useful tasks \cite{c.a1,c.a2,c.a3,c.a4,c.a5,c.a6,c.a7}. 
 In the resource theory of coherence, \textbf{the diagonal states} in a reference basis are chosen as the free states, which are known as incoherent states, and the incoherent operations that can not generate the coherence are known as free operations \cite{r.t4}.
So far, many efforts have been accomplished to understand this phenomenon and its relation with other quantum resources, such as quantum entanglement, quantum magic, quantum \textbf{discord, etc.} \cite{c.r1,c.r2,c.r3,c.r4,c.r5,c.r6,dis,qqc}.

The quantumness of open systems is extremely fragile due to the inevitable interactions with their \textbf{surrounding} environment, which leads to different noisy quantum channels \cite{nielsenbook}. It is very important to know the ability of these channels to change physical resources such as quantum coherence \cite{sk}, hence, the quantum resource theory can help us to understand how the quantum properties of a  system change under such evolutions. For this reason, in this paper, we investigate the coherence of quantum channels within the framework of the resource theory. To achieve this aim, we use Choi-Jamiolkowski isomorphism \cite{nielsenbook} and introduce \textbf{coherence-breaking} channels (CBCs) as free operations and their corresponding Choi states as free states \cite{c.breaking.c}. In our framework, the Quantum-incoherent relative entropy of coherence ($\mathcal{QI}$ REC) of quantum channels is equivalent to the $\mathcal{QI}$ REC of their corresponding Choi states \cite{c.r3,QI1}. We show that it consists of two parts: the relative entropy of coherence (REC) of open system and the basis-dependent quantum asymmetric discord \cite{qdiscord1,qdiscord2,qdiscord3}, where the former is zero for both quantum unital channels and quantum incoherent channels. On the other hand, one can prove the $\mathcal{QI}$ REC of channel is decreasing for divisible quantum incoherent channels and it can be a witness of non-Markovianity for quantum incoherent channels. \textbf{Also}, we demonstrate for qubit channels, the REC can be equivalent to the REC of their corresponding Choi states and the basis-dependent quantum symmetric discord can never be more than it. \textbf{Our results} will provide new light for a better understanding of the relationship between the coherence of channels and quantum correlations, and the coherence of a quantum channels can be considered as the quantumness of that.

We emphasize our formalism can be extended to the resource theory of entanglement for quantum channels. So, observing the analogical similarity between CBC and entanglement breaking channels (EBC) \cite{ee1}, one can define these operations as free channels and their corresponding Choi states as free states. In light of this, \textbf{one can} able to define the relative entropy of entanglement (REE) for quantum channels in the context of QRT \cite{ee2}.

The paper is organized as follows. In Sec. II,  a review of the quantum channels and the Choi representation is provided. In Sec. III, we briefly study the resource theory of coherence. We define the $\mathcal{QI}$ REC of quantum channels within the framework of the quantum resource theory in Sec. IV. We focus on
the coherence of qubit channels in Sec. V. \textbf{To illustrate} the coherence of channel, two examples are investigated in Sec. VI. The paper concludes in Sec. VII.
\section{Quantum channels}
A quantum channel is a linear map that satisfies the completely positive and \textbf{trace-preserving} (CPTP) conditions \cite{nielsenbook}. It can be shown that such mapping admits dilations as following form
\begin{equation}
\Lambda(\rho) =Tr_{E}[U_{SE}(\rho_{S}\otimes\rho_{E})U^{\dagger}_{SE}],
\end{equation}
where $\rho_{S}$ and $\rho_{E}$ are state of an open system and its environment, respectively, and  $U_{SE}$ denotes the unitary time evolution operator of the total system. On the other hand, the CPTP map $\Lambda$ can be represented by Kraus form \cite{nielsenbook}, and is shown by
\begin{equation}
\Lambda(\rho) =\sum_{i} K_{i}\rho K_{i},
\end{equation}
where $K_{i}$ are Kraus operators that $\sum  K^{\dagger}_{i} K_{i}=I$. Note that the quantum channels can have several Kraus representations.

In this work, we focus on another representation of the quantum channel where it is known as \textbf{the Choi matrix.} According to Choi-Jamiolkowski isomorphism, any CPTP map can be related to a density matrix of \textbf{the composite} system $AS$ in which $A$ is an auxiliary system with the same dimension $d$ as $S$. The Choi state of the channel $\Lambda$ is defined as
\begin{equation}
\Omega_{\Lambda}=(I_{A}\otimes\Lambda_{S})(\vert\Psi\rangle\langle\Psi\vert),
\end{equation}
where $\vert\Psi\rangle =\frac{1}{\sqrt{d}}\sum_{i} \vert ii\rangle$ is a maximally entangled state of $SA$. It is important to emphasize that there is a unique Choi state for every quantum channel includes all the channel's information \cite{nielsenbook}. Hence, we claim the coherence of any channel can be determined by the coherence of \textbf{its} corresponding Choi state.

\textbf{To define} the coherence of a channel, we review the resource theory of coherence in the next section.
\section{The resource theory of coherence}
As stated in the Introduction, quantum coherence is a physical property that is used as a resource for quantum systems. Hence, \textbf{the quantitative} determination of coherence for quantum systems has been extensively studied  \cite{c.m1,c.m2,c.m3,c.m4,c.m5}. Let us consider the Hilbert space $\mathcal{H}$ with fixed basis ${\vert i \rangle}_{i=0,...,d-1}$, then an incoherent state is defined as $\delta=\sum^{d-1}_{i}\delta_{i}\vert\ i\rangle\langle\ i\vert$.  The set of incoherent states is be denoted  by $\mathcal{I}$ and $\vert\psi\rangle=\frac{1}{\sqrt{d}}\sum^{d-1}_{i} e^{i \theta} \vert i\rangle$ is a maximally coherent state where $\theta_{i}$ is an arbitrary phase. A completely positive and trace preserving map $\Lambda$ is maximally incoherent (MIO) if $\Lambda(\delta) \in \mathcal{I}$ for any state $\delta \in \mathcal{I}$ \cite{mio}. Meanwhile, an incoherent operation (IO) has a Kraus representation such that  $\frac{K_{i}\delta K^{\dagger}_{i}}{Tr(K_{i}\delta K^{\dagger}_{i})} \in \mathcal{I}$ for all $n$ and $\delta \in\mathcal{I}$ \cite{r.t4}. By this restriction, the Kraus operators can be in the form $K_{i}=\sum^{d-1}_{j=0}c_{ij}\vert d_{i}(j)\rangle\langle j\vert$ for any incoherent operation, which are incoherent and $d_{i}(j)$ is a function of the index $j$ and $c_{ij}$ are coefficients \cite{inco}. If $d_{i}(j)$ is a permutation or one-to-one, then $K^{\dagger}_{i}$ is also incoherent as well as $K_{i}$. It is explicit from the definition that $IO\subseteq MIO$. In resource theory of coherence, the incoherent states and the incoherent operations are known as free states and free operations, respectively. A meaure for quantum coherence of state $\rho$ is characterized by a function $C(\rho)$ which satisfies the following properties \cite{r.t4}:
\begin{itemize}
\item[i]. $C(\rho)\geq 0$, for any $\rho$ and $C(\delta)=0$ if only if $\delta \in \mathcal{I}$;
\item[ii]. The coherence cannot increase under $MIO$ map $\Lambda$, i.e., $C(\Lambda(\rho))\leq C(\rho)$;
\item[iii]. For every $\Lambda\in IO$ with Kraus representation $\lbrace K_{i}\rbrace$, the coherence is non-increasing on average under selective measurement, i.e., $\sum_{i}p_{i}C(\rho_{i})\leq C(\rho)$, where $\rho_{i}=\frac{K_{i}\rho K^{\dagger}_{i}}{Tr(K_{i}\rho K^{\dagger}_{i})}$;
\item[iv]. The coherence cannot increase by mixing quantum states, i.e., $C(\sum_{i}p_{i}\rho_{i})\leq\sum_{i}p_{i}C(\rho_{i})$.
\end{itemize}

One of the measures that satisfy all the above requirements is the REC defined by \cite{r.t4}
 \begin{equation}\label{e4}
C_{r}(\rho)=\min_{\delta\in I}S(\rho\|\delta)=S(\rho^{d})-S(\rho),
\end{equation}
where $S(\rho\|\delta)=tr[\rho(Log{\rho}-Log{\delta})]$ is the relative entropy \cite{nielsenbook}, and the diagonal part of $\rho$ in the reference basis $\lbrace\vert i\rangle\rbrace$ is $\rho^{d}=\sum_{i}\vert\ i\rangle\langle\ i\vert\rho\vert\ i\rangle\langle\ i\vert$.

In the next section we use the resource theory and introduce the $\mathcal{QI}$ REC of channels. 

\section{The Resource theory of coherence for quantum channels}
In the previous section we introduced the resource theory of coherence for states. Here, we intend to define, in the context of the resource theory, a measure to determine the coherence of quantum channels. To achieve this purpose, we have to define the set of free states and free operations.

$\textbf{Free operations:}$
The free operations cannot generate \textbf{resources,} e.g., in resource theory of entanglement the LOCC are considered as free operations \cite{r.e1,r.e2,r.e3}. In this paper, we consider coherence-breaking channels $CBCs$ as free operations \cite{c.breaking.c}. A quantum incoherent channel $\Lambda$ is called coherence-breaking if $\Lambda(\rho)$ is an incoherent state for any state $\rho$. The set of all $CBCs$ denoted by $S_{cbc}$. \textbf{A coherence-breaking channel} kills any coherence present in the state and this is our motivation in this work.

$\textbf{Free states:}$ The Choi states corresponding to the free operations are regarded as the free states. The set of free states can be introduced by \textbf{the following form}
\begin{equation}
\mathcal{F}=\lbrace\Omega_{\Phi}\mid~\Phi\in S_{cbc}\rbrace.
\end{equation}
 The corresponding Choi states of all coherence-breaking channels have the following form \cite{c.breaking.c}
\begin{equation}
\Omega_{\Phi}=\sum_{i}\lambda_{i}\rho_{i}\otimes\vert i\rangle\langle i\vert.
\end{equation}

It is clear that $\mathcal{F}$ is a set of quantum-incoherent states and  $(\Lambda\otimes I) \Omega_{\Phi}\in\mathcal{F}$ for every state $\Omega_{\Phi}\in\mathcal{F}$ and $\Lambda\in{S_{cbc}}$. By using the Choi-Jamilkowski isomorphism, we define the $\mathcal{QI}$ REC of channels as \cite{c.r3,QI1}
\begin{equation}\label{e7}
C_{QI}(\Lambda)=C_{r}(\Omega^{A\vert S}_{\Lambda})=\min_{\Omega_{\Phi}\in\mathcal{F}}S(\Omega_{\Lambda}\|\Omega_{\Phi}),
\end{equation}
 with the minimization taken over the set of $\mathcal{F}$. Applying the Theorem $2$ in \cite{ref29}, $C_{QI}(\Lambda)$ can also be written as
 \begin{equation}
C_{QI}(\Lambda)=S(\Delta^{S}(\Omega_{\Lambda}))-S(\Omega_{\Lambda}),
\end{equation}
 in which $\Delta^{S}(\Omega_{\Lambda})=\sum_{i}(I\otimes \vert i\rangle\langle i\vert)\Omega_{\Lambda}(I\otimes \vert i\rangle\langle i\vert)$. Notice that the relative entropy of Choi states is monotonicity decreasing under the local quantum incoherent operations.
 
 In the following, we are going to obtain the relationship between the $\mathcal{QI}$ REC of channel and quantum correlations. For this purpose, we regard the basis-dependent quantum asymmetric discord \cite{qdiscord1,qdiscord2,qdiscord3}
 \begin{equation}\label{e9}
D^{A\vert S}(\Omega_{\Lambda})=I(\Omega_{\Lambda})-I[\Delta^{S}(\Omega_{\Lambda})],
\end{equation}
where $I(\rho_{AS})=S(\rho_{S})+S(\rho_{A})-S(\rho_{AS})$ is mutual information \cite{nielsenbook}. With the help of Eq.(\ref{e9}), it is straightforward to obtain the following equality
 \begin{equation}\label{e10}
C_{QI}(\Lambda)=C_{r}(\rho_{S})+D^{A\vert S}(\Omega_{\Lambda}).
\end{equation}
where $C_{QI}(\Lambda)$ compose of the REC of open system and the quantum asymmetric discord. \textbf{The Eq. (\ref{e10})} tells us that $C_{QI}(\Lambda)\geq D^{A\vert S}(\Omega_{\Lambda})$, or \textbf{in other words,} the quantum asymmetric \textbf{discord} can never exceed the $\mathcal{QI}$ REC for any quantum channels.

By taking partial trace over the ancilla $A$ one can obtain the state of the system $S$ as
\begin{equation}\label{e11}
\rho_{S}=Tr_{A}(\Omega_{\Lambda})=\frac{1}{d}\sum_{i}K_{i}K^{\dagger}_{i}.
\end{equation}
 Suppose $\Lambda$ is a quantum unital channel, i.e., $\sum_{i}K_{i}K^{\dagger}_{i}=\sum_{i}K^{\dagger}_{i}K_{i}=I$, so we conclude $C_r(\rho_{S})$ is zero for any quantum unital channel.

The state of the system,  Eq. (\ref{e11}) for an incoherent quantum channel has the following form
 \begin{equation}
\rho_{S}=\frac{1}{d}\sum^{d-1}_{l=0}\sum_{i}c_{il}c^{\ast}_{il}\vert d_{i}(l)\rangle\langle d_{i}(l)\vert,
\end{equation}
where the Kraus operator is  $K_{i}=\sum^{d-1}_{j=0}c_{ij}\vert d_{i}(j)\rangle\langle j\vert$.
Since $\rho_{S}$ in the above equation is an incoherent state then $C_{r}(\rho_{S})$ is also zero. So, the $\mathcal{QI}$ REC of the unital and the incoherent channels equals to  $D^{A\vert S}(\Omega_{\Lambda})$. Besides, we prove the $\mathcal{QI}$ REC is decreasing for divisible incoherent channels in the following proposition.

 $\emph{Proposition.}$ If $\Lambda$ is a divisible incoherent channel then $C_{QI}(\Lambda)$ is decreasing.
$\emph{Proof.}$  Assume $\Lambda$ is a divisible incoherent channel, i.e., $\Lambda_{t+\epsilon,0}=\Lambda_{t+\epsilon,t}\Lambda_{t,0}$. Using the facts that the $\mathcal{QI}$ REC in Eq.(\ref{e7}) is reduced under local incoherent channels, then we have $C^{A\vert S}_{r}(( I\otimes\Lambda)(\rho_{AS}))\leq C^{A\vert S}_{r}(\rho_{AS})$. So, one can write
\begin{eqnarray}\label{e13}
C^{A\vert S}_{r}(\rho_{AS}(t+\epsilon))&=&C^{A\vert S}_{r}((I\otimes\Lambda_{t+\epsilon,0})(\rho_{AS}(0))),\nonumber \\
&=&C^{A\vert S}_{r}((I\otimes\Lambda_{t+\epsilon,t}\Lambda_{t,0})(\rho_{AS}(0)))\nonumber \\
&=&C^{A\vert S}_{r}((I\otimes\Lambda_{t+\epsilon,t})(\rho_{AS}(t)))\nonumber \\
&\leq & C^{A\vert S}_{r}(\rho_{AS}(t)),
\end{eqnarray}
\textbf{The inequality (\ref{e13}) shows that} the coherence of the channel is decreasing and the proof is complete.\\
According to \textbf{the above} proposition the $\mathcal{QI}$ REC can be regarded as a witness for the non-Markovianity of quantum incoherent channels.

Now one can ask a question: what is the relation between the $\mathcal{QI}$ REC and the quantumness of a channel?
 Taking into account Eq.(\ref{e7}), it is clear as long as the $\mathcal{QI}$ REC of a channel is reduced, the channel will be closer to a $CBC$. This means that the coherence of \textbf{the output state of the channel} is also decreased, thus the output state becomes more classical. Therefore, the $\mathcal{QI}$ REC of channel can be considered as a measure of \textbf{the quantumness of the channel.} It is important to say that we interpret the quantumness, the amount of quantumness remains in \textbf{the state of the system} during the evolution, and this is different from the channel's ability to create quantum correlations.
\section{Coherence of qubit channels}
In this section, we consider the evolution of a single qubit under a quantum channel. An arbitrary qubit is expressed as $\rho=\frac{1}{2}(I+\vec{r}.\vec{\sigma})$ , where $\vec{r}$ is the $3$-dimensional Bloch vector with $\vec{r} \in R^{3}(\vert \vec{r} \vert\leq 1)$ and $\sigma_{i}$ are Pauli matrices.
A qubit channel, $\Lambda$, can be represented by a $4\times4$ matrix in the following form \cite{KI,RI}
\begin{equation}
\bf{F}= \left(
\begin{array}{cc}
1&\vec{0} \\
\vec{\tau} & T \\
\end{array}
\right),
\end{equation}
where, $T$ is a $3\times3$ real matrix and $\vec{\tau}$ and $\vec{0}$ are \textbf{3-dimensional} column and row vectors, respectively. Then we have
 \begin{equation}
\Lambda(\rho)=\frac{1}{2}[I+(T\vec{r}+\vec{\tau}).\vec{\sigma}].
\end{equation}

It is worthwhile to note that any qubit channel can be written as \cite{KI,RI}
 \begin{equation}
\bf{F}= \left(
\begin{array}{cccc}
1&0&0&0 \\
\tau^{\prime}_{1}& \lambda_{1}&0&0 \\
\tau^{\prime}_{2}&0&\lambda_{2}&0 \\
\tau^{\prime}_{3}&0&0&\lambda_{3}\\
\end{array}
\right),
\end{equation}
where, $\lambda$'s are singular values of the matrix $T$. For all coherence-breaking qubit channels the matrix $\bf{F}$ is in the following form \cite{c.breaking.c}
 \begin{equation}\label{e17}
\bf{F_{\Phi}}= \left(
\begin{array}{cccc}
1&0&0&0 \\
0&0&0&0 \\
0&0&0&0 \\
\tau_{3}&0&0&\lambda_{3}\\
\end{array}
\right),
\end{equation}
and the corresponding Choi matrix is
\begin{equation}
\Omega_{\Phi}= \left(
\begin{array}{cccc}
1+\tau_{3}-\lambda_{3}&0&0&0 \\
0&1+\tau_{3}+\lambda_{3}&0&0 \\
0&0&1-\tau_{3}+\lambda_{3}&0 \\
0&0&0&1-\tau_{3}-\lambda_{3}\\
\end{array}
\right).
\end{equation}

\textbf{Here,} $\Omega_{\Phi}$ is an incoherent matrix. Therefore, for qubit channels in the form of Eq. (\ref{e17}), the coherence of the channel coincides with the REC of its corresponding Choi state
\begin{equation}
C_{r}(\Lambda)=C_{r}(\Omega_{\Lambda})=\min_{\Omega_{\Phi}\in\mathcal{F}}S(\Omega_{\Lambda}\|\Omega_{\Phi}),
\end{equation}
and regarding the Eq.(\ref{e4}), we have
\begin{equation}\label{e20}
C_{r}(\Lambda)=S(\Omega^{d}_{\Lambda})-S(\Omega_{\Lambda}).
\end{equation}

Also, by using the basis-dependent quantum symmetric discord \cite{qdiscord1,qdiscord2,qdiscord3}
 \begin{equation}\label{e21}
D(\Omega_{\Lambda})=I(\Omega_{\Lambda})-I[(\Omega^{d}_{\Lambda})],
\end{equation}
The coherence of qubit channel gets the following form
\begin{equation}
C_{r}(\Lambda)=C_{r}(\rho_{S})+C_{r}(\rho_{A})+D(\Omega_{\Lambda}).
\end{equation}

 \textbf{The above} equation tells us that the quantum symmetric discord $D(\Omega_{\Lambda})$ \textbf{can never} exceed the coherence $C_{r}(\Lambda)$ for qubit channels.
\section{Examples}
In this section the coherence of \textbf{the channel} will be illustrated by means two examples.
\subsection{Amplitude damping channel}
Here, we calculate the coherence for an amplitude damping channel with Kraus operators $K_{0}=\vert 0\rangle\langle 0\vert+\sqrt{1-p}\vert 1\rangle\langle 1\vert$ and $K_{1}=\sqrt{p}\vert 0\rangle\langle 1\vert$ \cite{nielsenbook}. The corresponding Choi matrix for this channel is given by
\begin{equation}
\Omega_{\Lambda^{AD}}= \left(
\begin{array}{cccc}
\frac{1}{2}&0&0&\frac{\sqrt{1-p}}{2} \\
0& 0&0&0 \\
0&0&\frac{p}{2}&0 \\
\frac{\sqrt{1-p}}{2}&0&0&\frac{1-p}{2}\\
\end{array}
\right),
\end{equation}
 and the coherence of the channel will be
 \begin{equation}
C_{r}(\Lambda^{AD})=\frac{p-1}{2}\log{(\frac{1-p}{2})}+\frac{2-p}{2}\log{(\frac{2-p}{2})}+\frac{1}{2}.
\end{equation}

\begin{figure}
\centering 
\includegraphics[width=6cm]{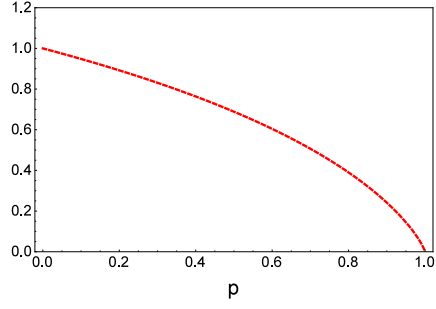}
\caption{(Color online). Plot of  $C_{QI}(\Lambda^{AD})$ as a function of the parameter $p$, for amplitude damping channel.}\label{case1}
\end{figure}

The behavior of the coherence of the channel (dashed red line) of the corresponding Choi state in terms of $p$ \textbf{is plotted} in Fig. \ref{case1}. From the figure one can see that the amount of coherence is reduced by increasing the parameter $p$ from 0 to 1 and it is very close to the quantumness $Q_{C}(\Lambda^{AD})$ which is defined in \cite{qqc}.
\subsection{Phase covariant channel}
Here, we consider a general model of qubit dynamics which includes dephasing, dissipation and heating effects \cite{ex1,ex2,ex}. The time-local master equation for this \textbf{evolution} is given by
\begin{eqnarray}\label{e25}
\frac{d\rho}{dt}&=&-i (\omega+h(t))[\sigma_{z},\rho]+\frac{\gamma_{z}(t)}{2}(\sigma_{z}\rho\sigma_{z}-\rho)\nonumber \\
&+&\frac{\gamma_{1}(t)}{2}(\sigma_{+}\rho\sigma_{-}-\frac{1}{2}\lbrace\sigma_{-}\sigma_{+}\rho\rbrace)\nonumber \\
&+&\frac{\gamma_{2}(t)}{2}(\sigma_{-}\rho\sigma_{+}-\frac{1}{2}\lbrace\sigma_{+}\sigma_{-},\rho\rbrace),\nonumber \\
\end{eqnarray}
where $\gamma_{i}(t) (i= 1,2,z)$ are \textbf{time-dependent} decay rates, $\sigma_{\pm}$ are the raising and lowering operators of the qubit, $\sigma_{z}$ is the Pauli spin operator in the $z$-direction and $h(t)$ time-dependent frequency shift and $\omega$ is the transition frequency of the qubit. It should be noted that the decay rates are time-dependent functions which can be \textbf{negative} at some times. The  Eq. (\ref{e25}) is the most general time-local master equations for a qubit which \textbf{indicates} a phase covariant transform. The Choi matrix for such a transformation will be in the following form
\begin{equation}
\small{\Omega_{\Lambda_{\omega}}= \left(
\begin{array}{cccc}
\frac{1+\kappa(t)+\eta _{\parallel}(t)}{4}&0&0& \frac{\eta_{\perp}(t)}{2} \\
0& \frac{1-\kappa(t)-\eta _{\parallel}(t)}{4}&0&0 \\
0&0& \frac{1+\kappa(t)-\eta _{\parallel}(t)}{4}&0 \\
\frac{\eta _{\perp}(t)}{2}&0&0&\frac{1-\kappa(t)+\eta _{\parallel}(t)}{4}\\
\end{array}
\right)},
\end{equation}
with 
\begin{eqnarray}
\kappa(t)&=&- e^{-\Gamma(t)}(1+2G(t))+1,\nonumber \\
\eta_{\parallel}(t)&=&e^{-\Gamma(t)},\nonumber \\
\eta_{\perp}(t)&=&e^{-\Gamma(t)/2-\Gamma_{z}(t)},\nonumber \\
\end{eqnarray}
where the terms in the above equations are defined as
\begin{eqnarray} 
\Gamma(t)&=&\int^{t}_{0}dt^{\prime}(\frac{\gamma_{1}(t^{\prime})}{2}+\frac{\gamma_{2}(t^{\prime})}{2}),\nonumber \\
\Gamma_{z}(t)&=&\int^{t}_{0}dt^{\prime}\gamma_{z}(t^{\prime}),\nonumber \\
G(t)&=&\int^{t}_{0}dt^{\prime}e^{\Gamma(t^{\prime})}\frac{\gamma_{2}(t^{\prime})}{2},\nonumber \\
\end{eqnarray}
It is clear from that $C_{r}(\rho_{S})=0$. The coherence of channel $\Lambda_{\omega}$ can be evaluated as
\begin{eqnarray}\label{e29}
C_{r}(\Lambda_{\omega})&=&-\frac{1\pm\kappa(t)+\eta_{\parallel}(t)}{4}\log{\frac{1\pm\kappa(t)+\eta_{\parallel}(t)}{4}}\nonumber \\
&+&\frac{1+\eta_{\parallel}(t)\pm\sqrt{\kappa^{2}(t)+4\eta_{\perp}^{2}(t)}}{4}
\times\log{\frac{1+\eta_{\parallel}(t)\pm\sqrt{\kappa^{2}(t)+4\eta_{\perp}^{2}(t)}}{4}}.\nonumber \\
\end{eqnarray}

Now let us assume both environment thermal and dephasing are at the same temperature $T$. Also, we ignore the effect of the  Lamb shift corrections of the first term. The decay rates of heating and dissipation reservior are $\gamma_{1}(t)/2=Nf(t)$ and $\gamma_{2}(t)/2=(N+1)f(t)$, respectively, where $N$ is the mean number of thermal photons. The function $f(t)$  depends on the form of the reservoir spectral density and  for a Lorentzian spectrum it is expressed as \cite{ex3}
\begin{equation}
f(t)=Re\lbrace\frac{\dot{c}(t)}{c(t)}\rbrace,
\end{equation}
with
\begin{equation}
c(t)=e^{-\frac{t}{2}}[\cosh{(\frac{dt}{2})}+\frac{1}{d}\sinh{(\frac{dt}{2})}]c(0),
\end{equation}
where $d=\sqrt{1-2R}$, and $R=\gamma_{0}/\lambda$ is a dimensionless positive number, in which $\gamma_{0}$ is an effective coupling constant and $\lambda$ is the width of the spectral density of the environment. For $R<1/2$ (weak coupling) the dynamics is divisible (Markovian) while for $R>1/2$ (strong coupling), it becomes non-divisible (non-Markovian).

 We consider the spectral density $J(\omega)=\alpha(\omega^{s}/\omega^{s-1}_{c})e^{-\omega/\omega_{c}}$  for pure dephasing dynamic, where $\omega_{c}$ is the cutoff frequency, $s$ the Ohmicity parameter and $\alpha$ the coupling constant. In this case, the decay rate for the dephasing channel is determined by \cite{ex4,ex5,ex6}
\begin{equation}
\gamma_{z}(t)=\int d\omega J(\omega)\coth{(\frac{\hbar\omega}{2k_{B}T})}\frac{\sin{(\omega t)}}{\omega}.
\end{equation}

The memory time of the dephasing environment can be defined by $1/\omega_{c}$. \textbf{To characterize} the relation between the cutoff frequency of the dephasing environment and the width of the spectral density of the thermal reservoir one can introduce a new parameter $\beta=\omega_{c}/\lambda$. In dephasing dynamics the non-Markovianity, $\gamma_{z}(t)<0$, occurs  whenever $s>s_{crit}(T=0) = 2$ \cite{ex7}. Hence, the dynamics of the whole system can be determined by two parameters $R$ and $s$.
 \begin{figure}[!h]
\begin{center}
\begin{tabular}{cc}
\includegraphics[width=5.5cm,height=4cm]{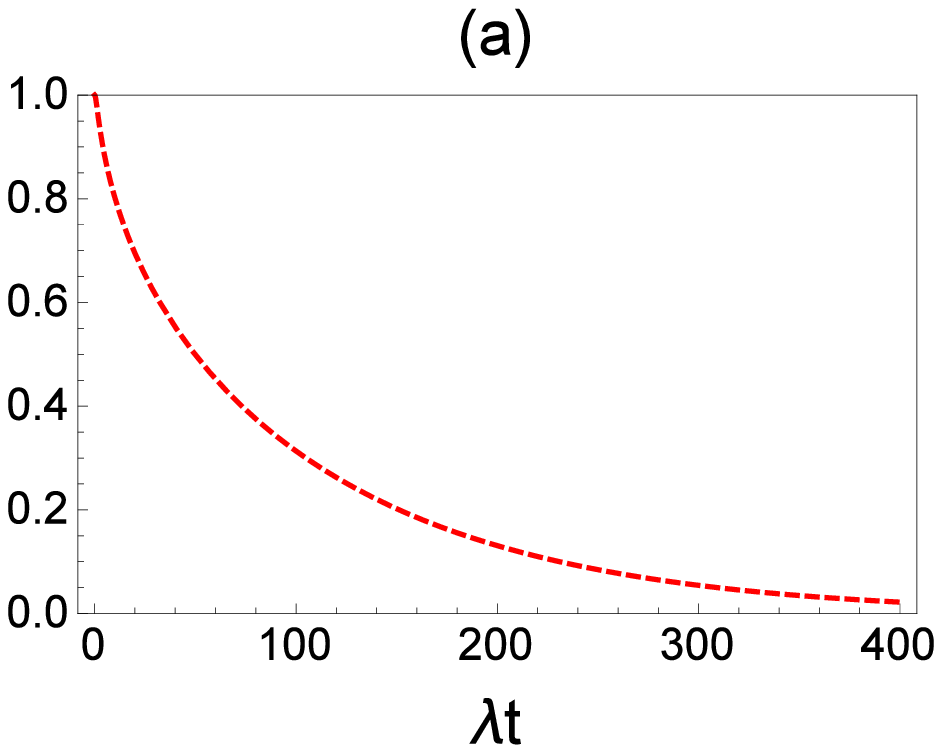}&\includegraphics[width=5.5cm,height=4cm]{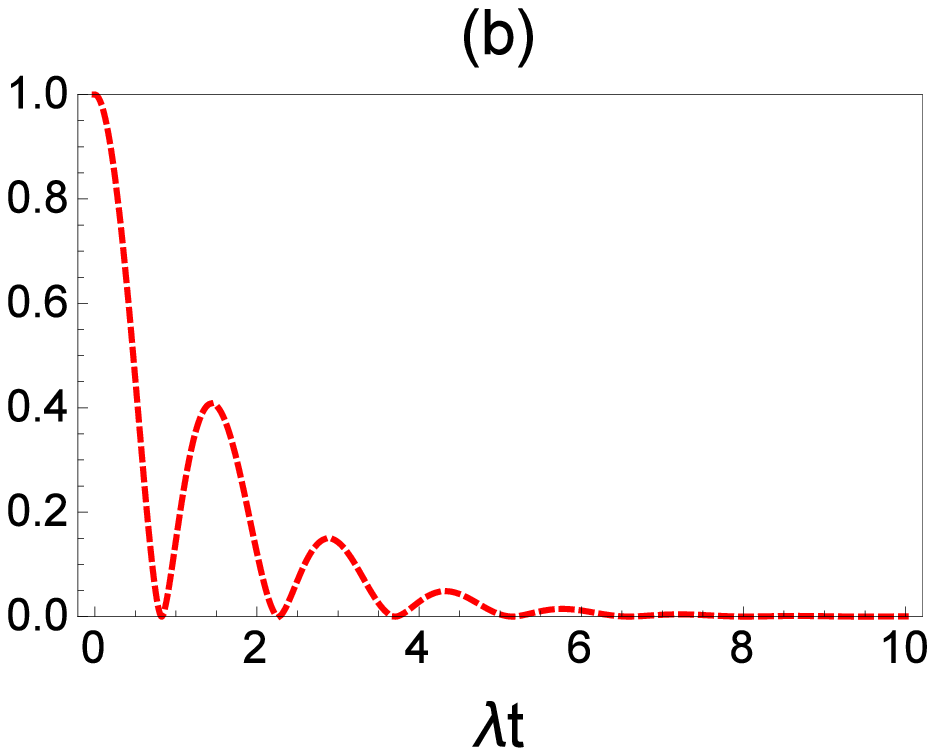}\\ 
\end{tabular}
\end{center}
\caption{(Color online). Dynamics of the coherence  $C_{r}(\Lambda_{\omega})$ as a function of $\lambda t$. (a)  the weak-coupling (Markovian) regime, $R=.01$  and $s=0.5$; (b) the strong-coupling (non-Markovian) regime, $R=10$  and $s=3.5$.}\label{case2}
\end{figure}

In this paper, we take $T=0$ thus the dephasing rate $\gamma_{z}(t)$ is independent of temperature and heating and dissipation rates are zero and $f(t)$, respectively. Then one can obtain the following expressions
\begin{eqnarray} 
&&\Gamma(t)=-\Re(\ln{[u(t)]}),\nonumber \\
&&\Gamma_{z}(t)=\frac{\alpha}{s-1}\widetilde{\Gamma}(s)(1-(1+\omega^{2}_{c}t^{2})\nonumber\\
&&\times [\cos{(s\arctan{(\omega_{c}t)})}+\omega_{c}t\sin{(s\arctan{(\omega_{c}t}))}]),\nonumber \\
&&\kappa(t)=-(1-\exp{[\Re(\ln{[u(t)]})]}),\nonumber \\
\end{eqnarray}
where $u(t)=\lbrace c(t)/c(0)\rbrace^{2}$ and $\widetilde{\Gamma}(s)$ is the Euler gamma function.

 At this point, we now return to the coherence problem of the phase covariant channel in Eq. (\ref{e29}). The dynamical behavior of coherence for quantum channel $\Lambda_{\omega}$ as a function of $\lambda t$ is shown in Fig.  \ref{case2}. We assume that the memory time of the dephasing and thermal environments is the same, i.e., we choose the parameter $\beta=1$. In Fig. \ref{case2}a, the evolution is Markovian since we have $s=0.5$ and $R=0.01$. The coherence of channel is ploted for $s=3.5$ and $R=10$ in Fig. \ref{case2}b, that the dynamics is non-Markovian. As can be seen the coherence of channel is decreasing for \textbf{the Markovian regime} over time while in non-Markovian regime the coherence of channel damply oscillates and increase in some time intervals, which this behavior is due to non-divisibility of the dynamics.\\
 
\section{Conclusion}                                                                                                                                                                        In this paper, we \textbf{have shown} the quantum-incoherent relative entropy of coherence ($\mathcal{QI}$ REC) of a quantum channel is equivalent to the $\mathcal{QI}$ REC of its Choi state. We have used the Choi-Jamiolkowski isomorphism, within the framework of QRT, to define it. Where, the coherence-breaking channels (CBCs) are considered as free operations and their corresponding Choi states as free states. It is demonstrated the $\mathcal{QI}$ REC consists of two parts: the REC of open system and the basis-dependent quantum asymmetric discord that the former is zero for both the quantum unital and quantum incoherent channels. Also, it is shown that the $\mathcal{QI}$ REC is decreasing for any divisible quantum incoherent channel and it can be considered as a witness of \textbf{the non-Markovianity} for incoherent channels. \textbf{Also, we have proposed} that the coherence of \textbf{the channel} can be regarded as the measure of the quantumness of that. Finally, we found that the coherence of qubit channels \textbf{can coincide} with the REC of their corresponding Choi states and the basis-dependent quantum symmetric discord can never exceed the coherence. Ultimately, it is worthwhile to note that our results could open a new way to better understand the relationship between quantum coherence as a physical resource and other quantum resources, such as quantum correlations.


\begin{thebibliography}{5}
\bibitem{r.t1} Brandao, F.G.S.L., Gour, G.: Reversible framework for quantum resource theories. Phys. Rev. Lett. {\bf 115}, 070503 (2015)
\bibitem{r.t2} Gour, G., Spekkens, R.W.: The resource theory of quantum reference frames: manipulations and monotones. New J. Phys. {\bf 10}, 033023 (2008)
\bibitem{r.t3} Brandao, F.G.S.L., Horodecki, M., Oppenheim, J., Renes, J. M., Spekkens,  R.W.: Resource theory of quantum states out of thermal equilibrium. Phys. Rev. Lett. {\bf 111}, 250404 (2013)
\bibitem{r.t4} Baumgratz, T., Cramer, M., Plenio, M.B.: Quantifying coherence. Phys. Rev. Lett. {\bf 113}, 140401 (2014)
\bibitem{r.e2} Horodecki, R., Horodecki, P., Horodecki, M., Horodecki, K.: Quantum entanglement. Rev. Mod. Phys.
{\bf 81}, 865–942 (2009)
\bibitem{r.e3} Plenio, M.B., Virmani, S.: An introduction to entanglement theory. Quantum Inf. Comput. {\bf 7}, 1 (2007)
\bibitem{as} Bartlett, S.D., Rudolph, T., Spekkens, R.W.: Reference frames, superselection rules, and quantum information. Rev. Mod. Phys. {\bf 79}, 555 (2007)
\bibitem{qt} Gour, G., Muller, M.P., Narasimhachar, V., Spekkens, R.W., Halpern,  N.Y.: The resource theory of informational nonequilibrium in thermodynamics. Phys. Rep. {\bf 583}, 1 (2015)
\bibitem{no} Brunner, N., Cavalcanti, D., Pironio, S., Scarani, V., Wehner, S.: Bell nonlocality. Rev. Mod. Phys. {\bf 86}, 419 (2014)
\bibitem{nom} Rivas, Á., Huelga, S.F., Plenio, M.B.: Quantum non-Markovianity: characterization, quantification and detection. Rep. Prog. Phys. {\bf 77}, 094001 (2014)
\bibitem{c.a1} Coles, P.J., Metodiev, E.M., L\"{u}tkenhaus, N.: Numerical approach for unstructured quantum key distribution. Nat. Commun.  {\bf 7}, 11712 (2016)
\bibitem{c.a2} Giovannetti, V., Lloyd, S., Maccone, L.: Advances in quantum metrology. Nat. Photonics {\bf 5}, 222 (2011)
\bibitem{c.a3} Yuan, X., Zhou, H., Cao, Z., Ma, X.: Intrinsic randomness as a measure of quantum coherence. Phys. Rev. A {\bf 92}, 022124 (2015)
\bibitem{c.a4} Streltsov, A., Adesso, G., Plenio, M.B.: Colloquium: Quantum coherence as a resource. Rev. Mod. Phys. \textbf{89}, 041003 (2017)
\bibitem{c.a5} Chin, S.: Coherence number as a discrete quantum resource. Phys. Rev. A {\bf 92}, 042336 (2017)
\bibitem{c.a6} Winter, A., Yang, D.: Operational resource theory of coherence. Phys. Rev. Lett. {\bf 116}, 120404 (2016)
\bibitem{c.a7} Chen, J.-J., Cui, J., Zhang, Y.-R., Fan, H.: Coherence susceptibility as a probe of quantum phase transitions. Phys. Rev. A {\bf 94}, 022112 (2016)
\bibitem{co} Horodecki, M., Oppenheim, J.: (Quantumness in the context of) resource theories. Int. J. Mod. Phys. B {\bf 27}, 1345019 (2013)
\bibitem{co1} Gour, G.: Quantum resource theories in the single-shot regime. Phys. Rev. A {\bf 95}, 062314 (2017)
\bibitem{co2} Liu, Z.-W., Hu, X., Lloyd, S.: Resource destroying maps. Phys. Rev. Lett. {\bf 118}, 060502 (2017)
\bibitem{co3} Sparaciari, C., del Rio, L., Scandolo, C.M., Faist, P., Oppenheim, J.: The first law of general quantum resource theories. arXiv:1806.04937 [quant-ph]
\bibitem{inco1} Chitambar, E., Gour, G.: Quantum resource theories. Phys. Rev. Mod. {\bf 91}, 025001 (2019)
\bibitem{c.r1} Chitambar, E., Hsieh, M.-H.: Relating the resource theories of entanglement and quantum coherence. Phys. Rev. Lett. {\bf 117}, 020402 (2016)
\bibitem{c.r2} Streltsov, A., Singh, U., Dhar, H.S., Bera, M.N., Adesso, G.: Measuring quantum coherence with entanglement. Phys. Rev. Lett. {\bf 115}, 020403 (2015)
\bibitem{c.r3} Chitambar, E., Streltsov, A., Rana, S., Bera, M.N., Adesso, G., Lewenstein, M.: Assisted distillation of quantum coherence. Phys. Rev. Lett. {\bf 116}, 070402 (2016).
\bibitem{c.r4} Xi, Z., Li, Y., Fan, H.: Quantum coherence and correlations in quantum system. Sci. Rep. {\bf 9}, 10922 (2015)
\bibitem{c.r5} Ma, J., Yadin, B., Girolami, D., Vedral,  V., Gu, M.: Converting coherence to quantum correlations. Phys. Rev. Lett. {\bf 116}, 160407 (2016)
\bibitem{c.r6} Mukhopadhyay, C., Sazim, S., Pati, A.K.: Coherence makes quantum systems 'magical'. Journal of Physics A: Mathematical and Theoretical {\bf 51 (41)}, 414006, (2018) 
\bibitem{dis} Wang, X.-L., Yue, Q.-L., Ya, C.-H., Gao, F., Qin, S.-J.: Relating quantum coherence and correlations with entropy-based measures. Sci. Rep. {\bf 7}, 12122 (2017)
\bibitem{qqc} Shahbeigi, F., Akhtarshenas, S. J.: Quantumness of quantum channels. Phys. Rev. A {\bf 98}, 042313 (2018)
\bibitem{nielsenbook} Nielsen, M.A., Chuang, I.L.: Quantum Computation and Quantum Information. Cambridge University Press, Cambridge (2000)
\bibitem{sk} Datta, C., Sazim, S., Pati, A.K., Agrawal, P.: Coherence of quantum channels. Annals of Physics, {\bf 397}, 243-258 (2018)
\bibitem{c.breaking.c} Bu, K., Swati, Singh, U., Wu, J.: Coherence-breaking channels and coherence sudden death. Phys. Rev. A {\bf 94}, 052335 (2016)
\bibitem{QI1} Streltsov, A., Rana, S., Bera, M.N., Lewenstein, M.: Towards resource theory of coherence in distributed scenarios. Phys. Rev. X {\bf 7}, 011024 (2017)
\bibitem{qdiscord1} Ollivier, H., Zurek, W.H.: Quantum discord: a measure of the quantumness of correlations. Phys. Rev. Lett. {\bf 88}, 017901 (2001)
\bibitem{qdiscord2} Luo, S.L.: Using measurement-induced disturbance to characterize correlations as classical or quantum. Phys. Rev. A {\bf 77}, 022301 (2008)
\bibitem{qdiscord3} Rulli, C.C., Sarandy, M.S.: Global quantum discord in multipartite systems. Phys. Rev. A {\bf 84}, 042109 (2011)
\bibitem{ee1} Horodecki, M., Shor, P.W., Ruskai,  M.B.: Entanglement breaking channels. Rev. Math. Phys. {\bf 15}, 629 (2003)
\bibitem{ee2} Vedral, V., Plenio, M.B.: Entanglement measures and purification procedures. Phys. Rev. A {\bf 57}, 1619 (1998)
\bibitem{c.m1} Shao, L.-H., Xi, Z., Fan, H., Li, Y.: Fidelity and trace-norm distances for quantifying coherence. Phys. Rev. A {\bf 91}, 042120 (2015)
\bibitem{c.m2} Bai, Z., Du, S.: Maximally coherent states. Quantum Inf. Comput. {\bf 15}, 1355 (2015)
\bibitem{c.m3} Mani, A., Karimipour, V.: Cohering and decohering power of quantum channels. Phys. Rev. A {\bf 92}, 032331 (2015)
\bibitem{c.m4} Du, S., Bai, Z., Qi, X.: Coherence measures and optimal conversion for coherent states. Quantum Inf. Comput. {\bf 15}, 1307 (2015)
\bibitem{c.m5} Yuan, X., Zhou, H., Cao, Z., Ma, X.: Intrinsic randomness as a measure of quantum coherence. Phys. Rev. A {\bf 92}, 022124 (2015)
\bibitem{mio} Aberg, J.: Quantifying superposition. arXiv:quant-ph/0612146
\bibitem{inco} Chitambar, E., Gour, G.: Comparison of incoherent operations and measures of coherence. Phys. Rev. A. {\bf 94}, 052336 (2016)
\bibitem{r.e1} Bennett, C.H., DiVincenzo, D.P., Smolin, J.A., Wootters, W.K.: Mixed-state entanglement and quantum error correction. Phys. Rev. A {\bf 54}, 3824 (1996)
\bibitem{ref29} Modi, K., Paterek, T., Son, W., Vedral, V., Williamson, M.: Unified view of quantum and classical correlations. Phys. Rev. Lett. {\bf 104}, 080501 (2010)
\bibitem{KI} King, C., Ruskai, M.B.: Minimal entropy of states emerging from noisy quantum channels IEEE Trans. Info. Theory. {\bf 47}, 192-209 (2001)
\bibitem{RI} Ruskai, M.B., Szarek, S., Werner, E.: An analysis of completely-positive trace-preserving maps on M2. Lin. Alg. Appl. {\bf 347}, 159 (2002)
\bibitem{m.i} Luo, S., Fu, S., Song, H.: Quantifying non-Markovianity via correlations Phys. Rev.  A {\bf 86}, 044101 (2012)
\bibitem{ex1} Smirne, A., Kolodynski, J., Huelga, S.F., Demkowicz-Dobrzanski, R.: Ultimate precision limits for noisy frequency estimation. Phys. Rev. Lett. {\bf 116} (12), 120801 (2016)
\bibitem{ex2} Lankinen, J., Lyyra, H., Sokolov, B., Teittinen, J., Ziaei, B., Maniscalco, S.: Complete positivity, finite-temperature effects, and additivity of noise for time-local qubit dynamics. Phys. Rev. A {\bf 93} (5), 052103 (2016)
\bibitem{ex} Tabesh, F.T., Karpat, G., Maniscalco, S., Salimi, S., Khorashad, A.S.: Time-invariant discord: high temperature limit and initial environmental correlations. Quantum Inf Process, {\bf 17}, 87 (2018)
\bibitem{ex3} Breuer, H.-P., Petruccione, F.: The Theory of Open Quantum Systems. Oxford University Press, Oxford (2002)
\bibitem{ex4} Luczka, J.: Spin in contact with thermostat: Exact reduced dynamics. Phys. A: Statistical Mechanics and its Applications, {\bf 167}, 919–934 (1990)
\bibitem{ex5} Palma, G.M., Suominen, K.-A., Ekert, A.-K.: Quantum computers and dissipation. Proc. Roy. Soc. London, Ser. A {\bf 452}, 567–584 (1996)
\bibitem{ex6} Reina, J.H., Quiroga, L., Johnson, N.F.: Decoherence of quantum registers. Phys. Rev. A {\bf 65}, 032326 (2002)
\bibitem{ex7} Haikka, P., Johnson, T.H., Maniscalco, S.: Non-Markovianity of local dephasing channels and time-invariant discord. Phys. Rev. A {\bf 87}, 010103(R) (2013)


\end{thebibliography}
\end{document}